\documentclass[aps,prl,showpacs,twocolumn,superscriptaddress,floatfix]{revtex4}

\usepackage{graphicx}
\usepackage{dcolumn}
\usepackage{epsfig}
\usepackage{bm}
\usepackage{latexsym}
\usepackage{amsfonts}
\usepackage{amssymb}
\usepackage{amsmath}
\usepackage{color}
\newcommand{\gp}{\dot{\gamma}}
\newcommand{\gpc}{\dot{\gamma}_c}
\newcommand{\gps}{\dot{\gamma}^\star}
\newcommand{\gpl}{\dot{\gamma}_{\hbox{\scriptsize\rm loc}}}
\newcommand{\gpt}{\dot{\gamma}_{\hbox{\scriptsize\rm eff}}}

\newcommand{\dd}{\hbox{\rm d}}

\begin{document}

\title{Shear-induced fragmentation of Laponite suspensions}

\author{Thomas Gibaud\footnote{Present address: Brandeis University, Physics Department, 415 South Street, Waltham, MA 02453, USA}}
\affiliation{Laboratoire de Physique, Universit\'e de Lyon -- \'Ecole Normale Sup\'erieure de Lyon  -- CNRS UMR 5672\\46 all\'ee d'Italie, 69364 Lyon cedex 07, France}
\author{Catherine Barentin}
\affiliation{Laboratoire de Physique de la Mati\`{e}re Condens\'{e}e et Nanostructures, Universit\'e de Lyon -- CNRS UMR 5586\\43 Boulevard du 11 Novembre 1918, 69622 Villeurbanne cedex, France}
\author{Nicolas Taberlet}
\affiliation{Laboratoire de Physique, Universit\'e de Lyon -- \'Ecole Normale Sup\'erieure de Lyon  -- CNRS UMR 5672\\46 all\'ee d'Italie, 69364 Lyon cedex 07, France}
\author{S\'{e}bastien Manneville}
\email{sebastien.manneville@ens-lyon.fr}
\affiliation{Laboratoire de Physique, Universit\'e de Lyon -- \'Ecole Normale Sup\'erieure de Lyon  -- CNRS UMR 5672\\46 all\'ee d'Italie, 69364 Lyon cedex 07, France}
\date{\today}

\begin{abstract}Simultaneous rheological and velocity profile measurements are performed in a smooth Couette geometry on Laponite suspensions seeded with glass microspheres and undergoing the shear-induced solid-to-fluid (or yielding) transition. Under these slippery boundary conditions, a rich temporal behaviour is uncovered, in which shear localization is observed at short times, that rapidly gives way to a highly heterogeneous flow characterized by intermittent switching from plug-like flow to linear velocity profiles. Such a temporal behaviour is linked to the fragmentation of the initially solid sample into blocks separated by fluidized regions. These solid pieces get progressively eroded over time scales ranging from a few minutes to several hours depending on the applied shear rate $\gp$. The steady-state is characterized by a homogeneous flow with almost negligible wall slip. The characteristic time scale for erosion is shown to diverge below some critical shear rate $\gps$ and to scale as $(\gp-\gps)^{-n}$ with $n\simeq 2$ above $\gps$. A tentative model for erosion is discussed together with open questions raised by the present results.
\end{abstract}
\pacs{83.60.La, 83.50.Rp, 83.60.Pq, 83.50.Lh}
\maketitle

\section{Introduction}
\label{s.intro}
Interest into the solid-to-fluid transition displayed by ``soft glassy materials'' (SGM) has risen tremendously in the past two decades not only because of its industrial importance but also due to its inherent theoretical issues and experimental challenges. On the theoretical side, the fact that this transition, also referred to as the ``jamming-unjamming'' transition, can be triggered either by increasing the temperature, by lowering the system concentration, or by applying a strong enough external load has led to propose a universal ``jamming phase diagram'' \cite{Liu:1998,Trappe:2001}. Although this picture remains appealing, it is clear that a complete understanding and modelling of glassy-like phenomena involved in materials with microstructures as diverse as submicron hard spheres, micronsized grains, short-range attractive particles, or highly charged platelets, are still out of reach \cite{Sciortino:2005}. Most recent experiments performed to investigate the structure and the dynamics of SGM at rest have relied on single or multiple light scattering techniques and have focused on {\it ageing} properties \cite{Cipelletti:2005} or on the presence of {\it dynamical heterogeneities} \cite{Weeks:2000,Duri:2006,Ballesta:2008}. Together, ageing and dynamical heterogeneities call for a spatiotemporal modelling of SGM that is still lacking.

Even more difficult is the task of elucidating the {\it flow mechanisms} in these out-of-equilibrium systems when submitted to some external shear. Rheology has long appeared as the only tool to probe the mechanical response of SGM and the shear-induced solid-to-fluid transition, also called the ``yielding transition.'' In particular, the problem of defining and measuring correctly the {\it yield stress} $\sigma_y$, i.e. the critical shear stress below which the material behaves as a solid and above which it flows like a liquid, has focused a lot of attention \cite{Barnes:1999}. The difficulties raised by the yield stress most often result from thixotropic features due to the competition between ageing and shear-induced ``rejuvenation'' of the sample \cite{Moller:2006}. Thus estimating precisely $\sigma_y$ involves applying or measuring very small shear rates over waiting times that can reach several hours.

\subsubsection*{Yield stress fluids and shear localization}

Some authors have proposed to distinguish between a ``static'' yield stress $\sigma_y^s$, defined for the material at rest, that would be the analogue of a static friction coefficient, and a smaller ``dynamic'' yield stress $\sigma_y^d$, measured on the flowing material by decreasing the imposed stress, similar to a dynamic friction coefficient \cite{Picard:2002}. Such a discrepancy linked to hysteresis effects also raises the question of whether the yielding transition in SGM is {\it continuous}, i.e. the shear rate $\gp$ continuously increases from zero when the shear stress $\sigma$ is increased above $\sigma_y$, or rather {\it discontinuous}, i.e. $\gp$ jumps to some finite critical value $\gpc$ as soon as $\sigma>\sigma_y$.

Continuous models for the flow curve $\sigma$ vs $\gp$ include the Bingham and the Herschel-Bulkley models, which have been recognized to hold for many SGM \cite{Barnes:1989,Sollich:1997}. On the other hand, a discontinuous yielding transition questions the basic assumption of homogeneous flow that underlies any standard rheological measurement \cite{Picard:2002,Picard:2005}. Indeed, if a shear rate $\gp<\gpc$ is applied, the flow is expected to display {\it shear localization}, i.e. the coexistence of a solid-like region where the local shear rate $\gpl$ vanishes and a fluid-like region where $\gpl=\gpc$.

The recent development of original experimental tools, such as active or passive microrheology \cite{Mason:1995a,Crocker:2000}, light scattering under shear, and time-resolved local velocimetry, that can go beyond standard rheology, has allowed one to address some of the above mentioned issues with renewed interest. In particular, using direct visualization, nuclear magnetic resonance (NMR), particle imaging, or ultrasonic velocimetry, shear localization was observed on a large variety of SGM, ranging from emulsions \cite{Coussot:2002,Bertola:2003,Becu:2006} or colloidal suspensions \cite{Pignon:1996,Raynaud:2002,Holmes:2004,Ragouilliaux:2006,Rogers:2008} to wet granular materials \cite{Barentin:2004,Huang:2005}. Still, in most of these experiments performed in Couette geometry, it remained unclear whether shear localization could be attributed to a truly non-zero $\gpc$ or to the stress inhomogeneity inherent to the concentric cylinder geometry. It is not until very recently that shear localization has been also evidenced in the cone-and-plate geometry where stress heterogeneity is minimized, thus establishing firmly the relevance of discontinuous models \cite{Moller:2008}.

\subsubsection*{Structure and rheology of Laponite suspensions}

Within the last decade, among the huge variety of SGM, Laponite, a synthetic clay of the hectorite type made of heavily charged disc-shaped particles of diameter 25--30~nm, thickness 1~nm, and density 2.5~g.cm$^{-3}$, has emerged as a good, albeit complicated, candidate to explore the above issues. When dispersed into water at a few weight percents, typically 0.6--4~wt.~\%, and depending on the ionic strength, Laponite suspensions evolve from a low-viscosity liquid to various solid-like ``arrested'' states. The sol--gel transition in Laponite has been investigated mostly using light scattering \cite{Mourchid:1995,Kroon:1996,Nicolai:2000,Ruzicka:2004,Ruzicka:2006}. The ageing properties of Laponite have triggered lots of effort and debate about the exact nature of these nonergodic states, in particular, about whether the material is actually a ``gel'' or a (Wigner) ``glass'' \cite{Ruzicka:2004,Mourchid:1998,Knaebel:2000,Tanaka:2004,Mongondry:2005,Jabbari-Farouji:2007b,Jabbari-Farouji:2008a,Ruzicka:2008}. Moreover, due to their out-of-equilibrium nature, Laponite suspensions were used to test possible violations of the fluctuation-dissipation theorem (FDT) \cite{Buisson:2003,Abou:2004}. Several recent microrheology experiments tend to prove that the effective temperature cannot be distinguished from the bath temperature, so that the FDT remains valid \cite{Jabbari-Farouji:2007a,Jop:2008pp}, but such results are still under debate \cite{Maggila:2009pp}.

The nonlinear rheology of Laponite suspensions has also been intensively studied with emphasis on thixotropy and ageing or rejuvenation under shear \cite{Mourchid:1995,Willenbacher:1996,Pignon:1997a,Pignon:1997b,Bonn:2002a,Abou:2003,Ianni:2007,Quemada:2008,Joshi:2008}. A few experiments have hinted to the presence of shear localization in the vicinity of the yield stress in Laponite samples \cite{Pignon:1996,Bonn:2002b,Ianni:2008}, which provided support for discontinuous models of the yielding transition \cite{Picard:2002,Moller:2008}. However, to the best of our knowledge, the local velocity field of Laponite suspensions was measured only in a wide-gap (2~cm) Couette cell with a temporal resolution of about 25~s per velocity profile using NMR velocimetry \cite{Bonn:2002b} and in a plate-plate geometry of gap 7~mm through dynamic light scattering (DLS) \cite{Ianni:2008}. The two corresponding papers only showed a couple of velocity profiles, the former raising the issue of strong geometry-induced stress heterogeneity and the latter reporting both wall slip and slow temporal oscillations of the velocity with the important limitation that DLS only provides point-like velocity measurements and necessitates to mechanically shift the cell in order to scan the whole gap. Therefore a systematic time-resolved study of velocity profiles in sheared Laponite suspensions would certainly provide new insights into the flow mechanisms involved in yielding.

\subsubsection*{Summary of our previous and present work}

In a previous work \cite{Gibaud:2008}, we have explored for the first time the influence of {\it boundary conditions} on yielding in Laponite samples by simultaneous rheology and time-resolved ultrasonic velocimetry. For the purpose of measuring local velocity using ultrasound, a significant 0.3~wt.~\% amount of hollow glass spheres of mean diameter 6~$\mu$m was added to our 3~wt.~\% Laponite suspensions. By comparing two experiments performed under similar imposed shear rates in smooth and rough (sand-blasted) Couette geometries on timescales of order 5000~s, we unveiled a dramatic effect of surface roughness on the flow mechanism during yielding. Indeed, while the scenario observed with rough walls was consistent with a discontinuous transition characterized by shear localization in agreement with previous observations on Laponite suspensions \cite{Pignon:1996,Bonn:2002b,Ianni:2008} and on other SGM \cite{Coussot:2002,Raynaud:2002,Ragouilliaux:2006,Rogers:2008,Moller:2008}, slippery boundary conditions in the smooth cell led to a thoroughly different picture. When wall slip was allowed, the sample was reported to break up into macroscopic solid pieces that are slowly eroded by the surrounding fluidized material up to the point where the whole sample has become fluid.

The aim of the present paper is to provide a full data set, more discussion, and a toy model of this original yielding scenario triggered by slippery boundary conditions in Laponite suspensions seeded with microspheres. Our paper is organized as follows. Section~\ref{s.exper} describes the materials and methods used to study the yielding transition in Laponite. Results obtained in a smooth Couette cell on a 3~wt.~\% Laponite sample seeded with 1~wt.~\% microspheres are presented in Section~\ref{s.results} for a large range of applied shear rates. Finally, Section~\ref{s.discuss} provides a discussion of the results, in particular about the possible influence of the microsphere concentration, and an attempt to model the experimental observations.

\section{Experimental}
\label{s.exper}

\subsection{Sample preparation}
\label{s.sample}
Laponite powder (Rockwood, grade RD) is used as received and dispersed at 3~wt.~\% in ultrapure water with no added salt. The pH of the solution is approximately 10. According to the most recent litterature \cite{Jabbari-Farouji:2008a,Ruzicka:2008}, such a suspension (3~wt.~\% Laponite without salt) is supposed to fall into the Wigner glass region of the phase diagram.

In the present study, the Laponite suspensions are immediately seeded with hollow glass spheres of mean diameter 6~$\mu$m (Sphericel, Potters) at a weight fraction $0.3$ or 1~wt.~\%. As explained below in Section~\ref{s.setup}, these microspheres act as contrast agents for ultrasonic velocimetry. Most of current protocols used for Laponite preparation recommend to filter the samples with a 0.4--0.8~$\mu$m mesh size after vigorous stirring for 15--30~min. Such filtration is believed to break up Laponite clusters that may form due to incomplete dissolution and may lead to erroneous interpretations of the sample structure as ``gel-like'' from light scattering data \cite{Bonn:1999}. Here, however, due to the presence of the microspheres, subsequent filtration of the samples is not possible.

Within 30~min of magnetic stirring, the dispersion becomes homogeneous and very viscous but remains fluid. Due to the presence of the microspheres, the solution is slightly turbid, which will allow for direct visual inspection of the sheared samples. The sample is then left to rest at room temperature. After about two hours, the sample viscosity has increased by several orders of magnitude and the dispersion has clearly built up a yield stress and solid-like properties since it is able to sustain its own weight. We let ageing proceed for at least two days, i.e. for waiting times $t_w\gtrsim 2\;10^5$~s. For such large $t_w$, we expect the ageing dynamics to have slowed down so much that the influence of ageing can be neglected on the timescales of typically one hour involved in the present experiments. In any case, as discussed below, the samples are pre-sheared before any experiments in order to erase most of the sample history through shear rejuvenation.

\subsection{Rheological protocol}
\label{s.rheo_protocol}
\begin{figure}[ht]
\begin{center}
\includegraphics{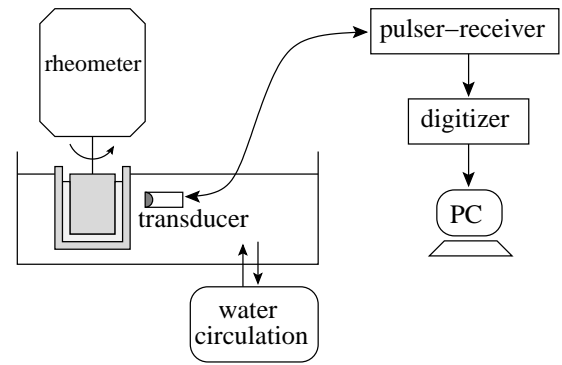}
\caption{Sketch of the experimental setup for combined rheological and velocity profile measurements. The Couette cell is shown in gray.}
\label{f.setup}
\end{center}
\end{figure}

Linear and nonlinear rheological properties are measured using a stress-controlled rheometer (Bohlin C-VOR 150, Malvern Instruments) in Mooney-Couette (concentric cylinder) geometry. Our cell is made out of Plexiglas without further treatment. The bob (inner cylinder) of radius $R_1=24$~mm is rotating and will be called the ``rotor'' in the following. It is terminated by a cone with an angle of 2.3$^\circ$. The stator, i.e. the fixed cup (outer cylinder), has a radius $R_2=25$~mm. The gap width of this Mooney-Couette geometry is thus $e=R_2-R_1=1$~mm. The curvature of our cell leads to a stress decrease of about 8~\% from the rotor to the stator. The height of the rotor is $H=30$~mm. The standard deviation of height profiles of the cylindrical walls measured using atomic force microscopy is typically 15~nm, which will be referred to as  ``smooth.'' The whole cell is immersed in a water tank of volume of about 1~L connected to a water bath whose temperature is kept constant and equal to 25$\pm 0.1^\circ$C (see Fig.~\ref{f.setup}).

Before any measurement, the sample is {\it pre-sheared} for 1~min at +1500~s$^{-1}$ and for 1~min at -1500~s$^{-1}$. Such a protocol erases most of the sample history through shear rejuvenation and ensures that the strain of the sample accumulated during loading into the cell has no influence  \cite{Willenbacher:1996}. We then proceed with a standard oscillatory test at 1~Hz in the linear regime for 2~min. This test allows us to make sure that the values of the viscoelastic moduli $G'\simeq 500$~Pa and $G''\simeq 25$~Pa no longer change significantly at the beginning of the actual experiment. We also checked that this procedure leads to reproducible results over a few hours. Finally, at time $t=0$, the experiment proceeds either with the standard (linear or nonlinear) rheological tests shown in Section~\ref{s.rheo} or with combined rheological and velocimetry measurements (see Section~\ref{s.usv}). In the following, $\sigma$ and $\gp$ denote the shear stress and the shear rate indicated by the rheometer. The rotor velocity $v_0$ and $\gp$ are linked by:
\begin{equation}
v_0=\frac{R_1(R_1+R_2)}{R_1^2+R_2^2}\,\gp e\,,
\label{e.v0}
\end{equation}
where the geometrical factor $R_1(R_1+R_2)/(R_1^2+R_2^2)$ accounts for the cell curvature \cite{Salmon:2003d}. In the presence of wall slip or heterogeneous flows, $\gp$, the so-called ``engineering'' or ``global'' shear rate may strongly differ from the local shear rate which will be noted $\gpl$ or $\gp(r)$, where $r$ is the radial distance from the rotor.

\subsection{Ultrasonic velocimetry and optical imaging}
\label{s.setup}
Our setup for combined rheology and local velocimetry is sketched in Fig.~\ref{f.setup}. The sample velocity field is measured using ultrasonic speckle velocimetry (USV) at about 15 mm from the cell bottom. USV is a technique that allows one to access velocity profiles in Couette geometry with a spatial resolution of 40~$\mu$m and a temporal resolution of 0.02--2~s depending on the shear rate. It relies on the analysis of successive ultrasonic speckle signals that result from the interferences of the backscattered echoes of successive incident pulses of central frequency 36~MHz generated by a high-frequency piezo-polymer transducer (Panametrics PI50-2) connected to a broadband pulser-receiver (Panametrics 5900PR with 200 MHz bandwidth). The speckle signals are sent to a high-speed digitizer (Acqiris DP235 with 500 MHz sampling frequency) and stored on a PC for post-processing using a cross-correlation algorithm that yields the local displacement from one pulse to another as a function of the radial position $r$ across the gap. One velocity profile is then obtained by averaging over typically 1000 successive cross-correlations. Full details about the USV technique may be found in Ref.~\cite{Manneville:2004a}.

The scattered signal is provided either by the material microstructure itself or by seeding the fluid with ``contrast agents'' with the constraint to remain in the single scattering regime. In the case of Laponite samples, which are transparent to ultrasound, we used the microspheres described above to produce ultrasonic echoes in a controlled way. The sound speed in our samples was independently measured to be $c=1495\pm 5$~m.s$^{-1}$ at 25$^\circ$C.

The microspheres also lead to a slight turbidity which allows for a direct visualization of the sheared samples. A simple CCD camera (Cohu 4192) is set in front of the water tank and the Couette cell is lit from behind. Images of the samples are recorded at a frame rate of 5~fps and later synchronized with the velocity profile measurements as explained in our previous work \cite{Gibaud:2008}. In this paper, we shall only show images typical of the various stages of the experiment.

\section{Results}
\label{s.results}

\subsection{Conventional rheometry}
\label{s.rheo}
\begin{figure}[ht]
\begin{center}
\includegraphics{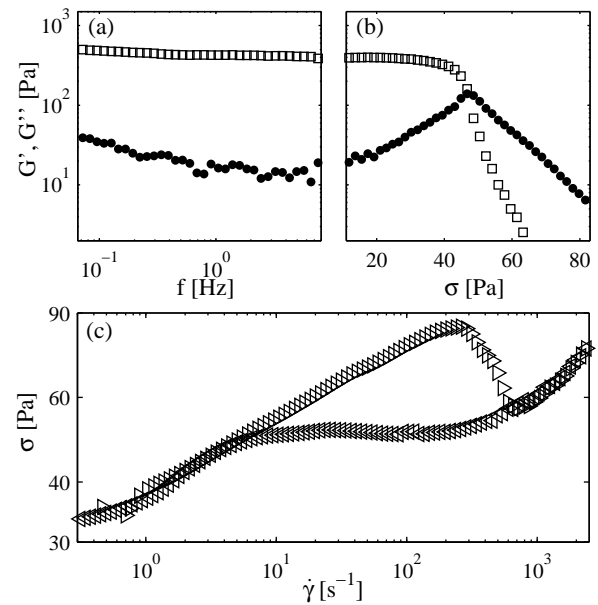}
\caption{Rheology of a 3~wt.~\% Laponite suspension seeded with 1~wt.~\% microspheres. Viscoelastic moduli $G'$ ($\square$) and $G''$ ($\bullet$) as a function of (a) oscillation frequency for a fixed shear stress amplitude of 0.5~Pa and (b) shear stress amplitude for a fixed frequency of 1~Hz. (c) Flow curve $\sigma$ vs $\gp$ obtained by sweeping up ($\triangleright$) the shear rate for 200~s and down ($\triangleleft$) for 200~s. All measurements are performed on fresh samples from the same batch using the protocol described in Section~\ref{s.rheo_protocol}.}
\label{f.rheo}
\end{center}
\end{figure}

Standard rheological measurements performed on a 3~wt.~\% Laponite suspension seeded with 1~wt.~\% microspheres are shown in Fig.~\ref{f.rheo}. The viscoelastic moduli displayed in Fig.~\ref{f.rheo}(a) and measured in the linear regime for a shear stress amplitude of 5~Pa (which corresponds to a strain of at most 0.13~\% over the accessible range of frequencies) are typical of an arrested (gel or glass) state. The elastic modulus $G'$ is always much larger than the loss modulus $G''$ and only decays from 500~Pa to 400~Pa over frequencies $f=0.07$--8~Hz. $G''$ also slowly decreases from about 30~Pa to 15~Pa over this range of frequencies and shows no sign of downturn at low frequencies, a feature that is usually attributed to the presence of very slow relaxation modes in SGM \cite{Sollich:1997,Mason:1995a}.

Figure~\ref{f.rheo}(b) shows the evolution of the viscoelastic moduli in the nonlinear regime for shear stress amplitudes above 15~Pa at a fixed frequency of 1~Hz. These data clearly point to a yield stress of 47~Pa, when defined as the shear stress amplitude $\sigma_y$ for which $G'(\sigma_y)=G''(\sigma_y)$. This yield stress corresponds to a strain of 25~\%. Above $\sigma_y$, $G'$ drops dramatically and rapidly becomes negligible when compared to $G''$. In other words, the sample becomes {\it fluid} above $\sigma_y$.

All these measurements are fully consistent with previous data on Laponite suspensions \cite{Mourchid:1998,Willenbacher:1996,Cocard:2000}. Note that the addition of microspheres influences the ionic strength of the suspension. The conductivity of a 1~\% wt. suspension of our glass microspheres in water was measured to be about 100~$\mu$S.cm$^{-1}$, which is roughly equivalent to a salt (NaCl) concentration of $0.8\;10^{-3}$~mol.L$^{-1}$. Increasing the microsphere concentration induces a noticeable stiffening of our samples: $G'$ increases by about 20~\% when the microsphere concentration is increased from 0.3~wt.~\% to 1~wt.~\% probably due to the change in ionic strength \cite{Mourchid:1998}.

As explained in the introduction, another way to probe the yielding transition using rheology is to perform nonlinear measurements where a constant shear stress or shear rate is imposed. Since the flow curve is almost flat for small shear rates, it is usually more suitable to work under imposed shear rate. Figure~\ref{f.rheo}(c) shows the flow curve obtained through a shear rate sweep from 0.3 to 2500~s$^{-1}$ within 200~s followed by the corresponding downward sweep at the same rate. The most obvious feature of this flow curve is the large hysteresis between upward and downward sweeps. This is typical of {\it thixotropic} materials whose microstructure is slowly modified by shear resulting in a time-dependent apparent viscosity $\eta=\sigma/\gp$. The very same kind of hysteresis cycle was observed recently for similar shear rates in a drilling mud \cite{Ragouilliaux:2006}.

Following previous works \cite{Bertola:2003,Ragouilliaux:2006}, we anticipate that the yield stress corresponds to the value of the shear stress on the plateau observed during the downward sweep. This yields $\sigma_y\simeq 51$~Pa in satisfactory agreement with the previous estimate of 47~Pa, so that one has $\sigma_y=49\pm 2$~Pa. Finally, the shape of the flow curve at low shear rates ($\gp\lesssim 0.5$~s$^{-1}$) points to significant wall slip which is highly probable in our smooth geometry \cite{Bertola:2003}. However, as we shall see below through time-resolved measurement of the local velocity, interpreting such a non-stationary flow curve is difficult due to complex temporal behaviours that take place over time scales of about one hour.

\subsection{Combined velocimetry and rheology}
\label{s.usv}
In this section, we present the combined rheological and velocity profile measurements performed under imposed shear rate on a 3~wt.~\% Laponite sample seeded with 1~wt.~\% microspheres in a smooth Couette cell. As recalled in the introduction, our previous experiments \cite{Gibaud:2008} with a lower microsphere concentration of 0.3~wt.~\% showed that slippery boundary conditions may lead to a rather complex scenario of slow fragmentation and erosion of the initially solid material. In paragraph \ref{s.results.sr}, we shall explore the influence of the imposed shear rate on this scenario for a given microsphere concentration of 1~wt.~\% through experiments performed over typically 5000~s. The influence of the microsphere concentration will be addressed in paragraph \ref{s.results.conc}.

In paragraph \ref{s.usv.early} below, we focus on the velocity profiles recorded within the first few mi\-nu\-tes after the inception of shear. We show that the flow may first seem compatible with a simple shear localization picture of critical shear rate $\gpc\simeq 125$~s$^{-1}$. However, as explained later in paragraph \ref{s.usv.long}, shear localization does not correspond to the stationary state reached by the system and the experiments have to be conducted over time scales of the order of one hour to probe the actual long-time flow behaviour. 

\subsubsection{Early stage after shear start-up}
\label{s.usv.early}
\begin{figure}[ht]
\begin{center}
\includegraphics{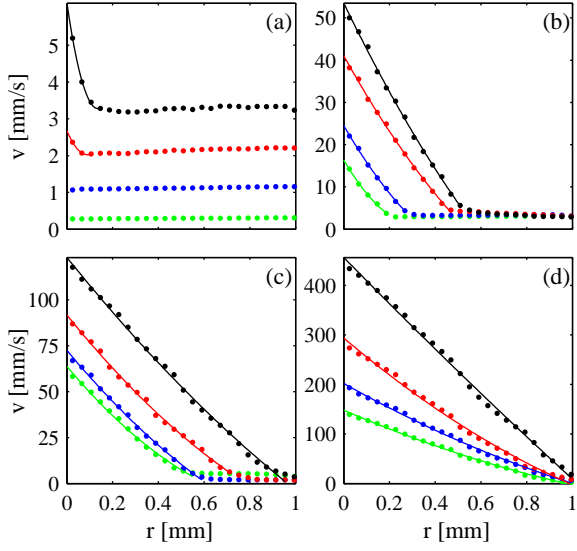}
\caption{Velocity profiles recorded during the early stage of our experiments on a 3~wt.~\% Laponite suspension seeded with 1~wt.~\% microspheres for imposed shear rates $\gp=$ (a) 1 (green), 5 (blue),  11 (red), 20 (black), (b) 30 (green), 40 (blue), 50 (red), 60 (black), (c) 70 (green), 80 (blue), 100 (red), 125 (black), (d) 150 (green), 200 (blue), 300 (red), and 500~s$^{-1}$ (black). $r$ is the radial distance from the rotor. The velocity profiles were averaged over $t=30$--40~s. The solid lines show quadratic fits of the data performed over the flowing region.}
\label{f.early}
\end{center}
\end{figure}

Figure~\ref{f.early} displays the velocity profiles $v(r)$ averaged over $t=30$--40~s after various shear rates $\gp$ ranging from 1 to 500~s$^{-1}$ are applied at time $t=0$. The time window for averaging the velocity profiles was chosen so that fast initial transients due to flow inception have died out. Such transients last typically a few seconds, after which the dynamics becomes much slower and the velocity profiles do not significantly change over 10~s. In the following, we shall analyze these velocity profiles as if they corresponded to steady measurements, keeping in mind that, over time scales longer than a few minutes, a much more complex behaviour will emerge.

For $\gp\lesssim 10$~s$^{-1}$, total wall slip is observed and the material undergoes solid-body rotation. In this case, the shear rate effectively experienced by the material vanishes, a feature that was already mentioned in Ref.~\cite{Ianni:2008}. Above $\gp=10$~s$^{-1}$, the velocity profiles are characterized by a flowing zone close to the rotor that coexists with an ``arrested'' solid-like region where the local shear rate is zero. The size of the fluid-like region increases with the imposed shear rate. For $\gp\gtrsim 125$~s$^{-1}$, velocity profiles are homogeneous and the whole sample is fluid on the time window investigated here, i.e. after a few 10~s. These measurements are reminiscent of shear localization as reported in various other SGM \cite{Coussot:2002,Ragouilliaux:2006,Rogers:2008,Moller:2008}. 

Note that apparent wall slip remains significant at both walls as long as some ``arrested'' region is present. Indeed, for $\gp\lesssim 125$~s$^{-1}$, the velocity of the sample in the close vicinity of the cell boundaries never reaches that of the walls, i.e. $v(r=0)<v_0$ at the rotor (where $v_0$ is given by Eq.~(\ref{e.v0})) and $v(r=e)>0$ at the stator. Wall slip becomes negligible only when a homogeneous flow is recovered for $\gp\gtrsim 125$~s$^{-1}$. In the following, in order to compare our data with predictions and experiments obtained in the absence of wall slip, we shall focus on the effective global shear rate $\gpt$ defined by \cite{Salmon:2003d}:
\begin{equation}
\gpt=\frac{R_1^2+R_2^2}{R_1(R_1+R_2)}\,\frac{v(0)-v(e)}{e}\,,
\end{equation}
rather than on the applied shear rate $\gp$.

\begin{figure}[ht]
\begin{center}
\includegraphics{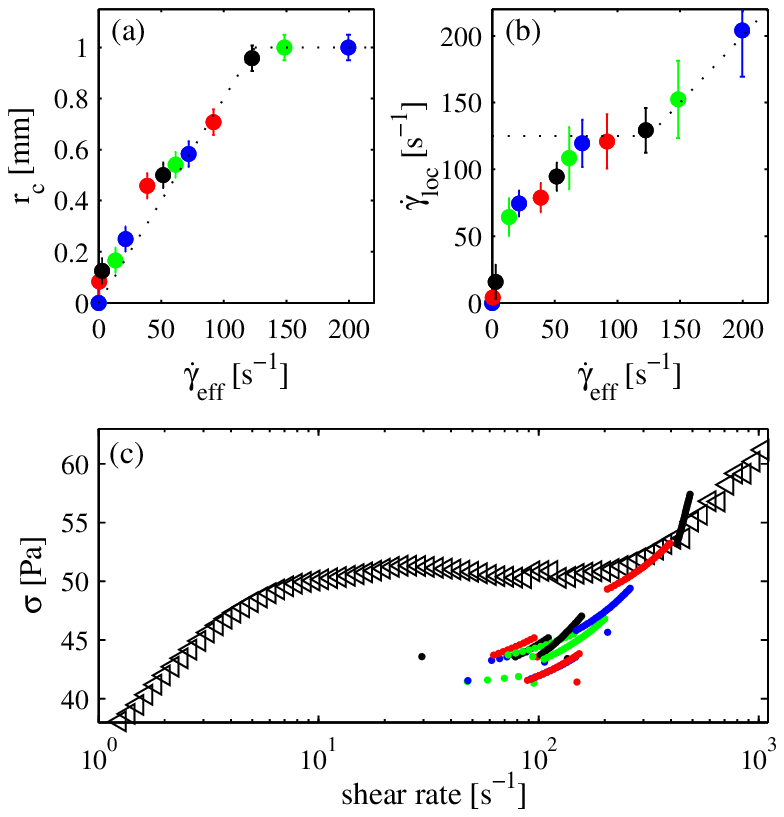}
\caption{(a) Width $r_c$ and (b) local shear rate $\gpl$ of the fluid-like region as a function of the effective global shear rate $\gpt$. (c) Local rheological flow curves $\sigma(r)$ vs $\gp(r)$ (colors) together with the global rheological data recorded during the downward sweep shown in Fig.~\ref{f.rheo} ($\triangleleft$). The colors correspond to the various velocity profiles of Fig.~\ref{f.early} (see text). The dotted lines in (a) and (b) indicate shear localization with a constant $\gpc=125$~s$^{-1}$.}
\label{f.locrheol}
\end{center}
\end{figure}

By using quadratic fits of the velocity profiles within the fluid-like band, one may easily extract the width $r_c$ of the flowing zone as well as the local shear rate $\gpl$ averaged over this sheared region. Such an analysis is presented as a function of the effective global shear rate $\gpt$ in Fig.~\ref{f.locrheol}(a) and (b). The error bars in Fig.~\ref{f.locrheol}(b) correspond to the standard deviation of the fitted local shear rate. It can be seen that the proportion of the sheared region increases roughly linearly with $\gpt$, consistently with the prediction of the simplest theoretical scenario for shear localization \cite{Picard:2002,Coussot:2002} and with recent experimental findings on a colloidal gel \cite{Moller:2008}. Here, Fig.~\ref{f.locrheol}(a) clearly points to a critical shear rate $\gpc\simeq 125$~s$^{-1}$.

Yet, Fig.~\ref{f.locrheol}(b) contradicts such a simple shear localization scenario. Indeed, it is clearly seen that the local shear rate $\gpl$ is {\it not} constant and equal to $\gpc$. It rather increases sharply from $\gpl\simeq 60$~s$^{-1}$ to $125$~s$^{-1}$ as $\gp$ is increased. This is confirmed by looking at the ``local'' flow curves displayed in Fig.~\ref{f.locrheol}(c) and showing the local shear stress $\sigma(r)$ plotted against the local shear rate $\gp(r)$. $\sigma(r)$ is computed from the global shear stress $\overline{\sigma}$ measured by the rheometer simultaneously to the velocity profiles using \cite{Salmon:2003d}:
\begin{equation}
\sigma(r)=\frac{2 R_1^2 R_2^2}{(R_1^2+R_2^2) r^2}\,\overline{\sigma}\,,
\end{equation}
where the proportionality factor accounts for the Couette geometry. The local shear rate $\gp(r)=-r\frac{\partial}{\partial r}\left(\frac{v}{r}\right)$ is simply estimated from the derivative of the quadratic fits of the velocity profiles of Fig.~\ref{f.early}. The resulting local flow curves $\sigma(r)$ vs $\gp(r)$ are compared to the downward sweep of Fig.~\ref{f.rheo}(c) in Fig.~\ref{f.locrheol}(c). For the highest shear rates, the local data collapse rather well on the global flow curve. This indicates that for $\gp\gtrsim 200$~s$^{-1}$, the material becomes fully fluid-like almost instantly and that no time-dependent phenomena further occur. For smaller shear rates, the local flow curves reveal two interesting features: (i) the existence of a ``forbidden'' range of local shear rates since $\gp(r)\lesssim 60$~s$^{-1}$ is never observed and (ii) the relevance of time-dependent, history-dependent, or metastable phenomena since the local shear stress for $t=30$--40~s is significantly smaller than the shear stress recorded during the downward sweep of Fig.~\ref{f.rheo}(c). Finally, for shear-localized velocity profiles, $\gp(r)$ covers the range 60--125~s$^{-1}$.

In the next paragraph, we shall see that shear localization as revealed in Fig.~\ref{f.early} is {\it only transient} so that the non-standard scenario shown in Fig.~\ref{f.locrheol}(b) and (c) is not so surprising. In any case, it is important to stress the fact that, if the velocity measurements had been stopped after a few minutes, shear localization could have been wrongly interpreted as the steady-state for yielding in this SGM.

\subsubsection{Long-time flow behaviour}
\label{s.usv.long}
\begin{figure}[ht]
\begin{center}
\includegraphics{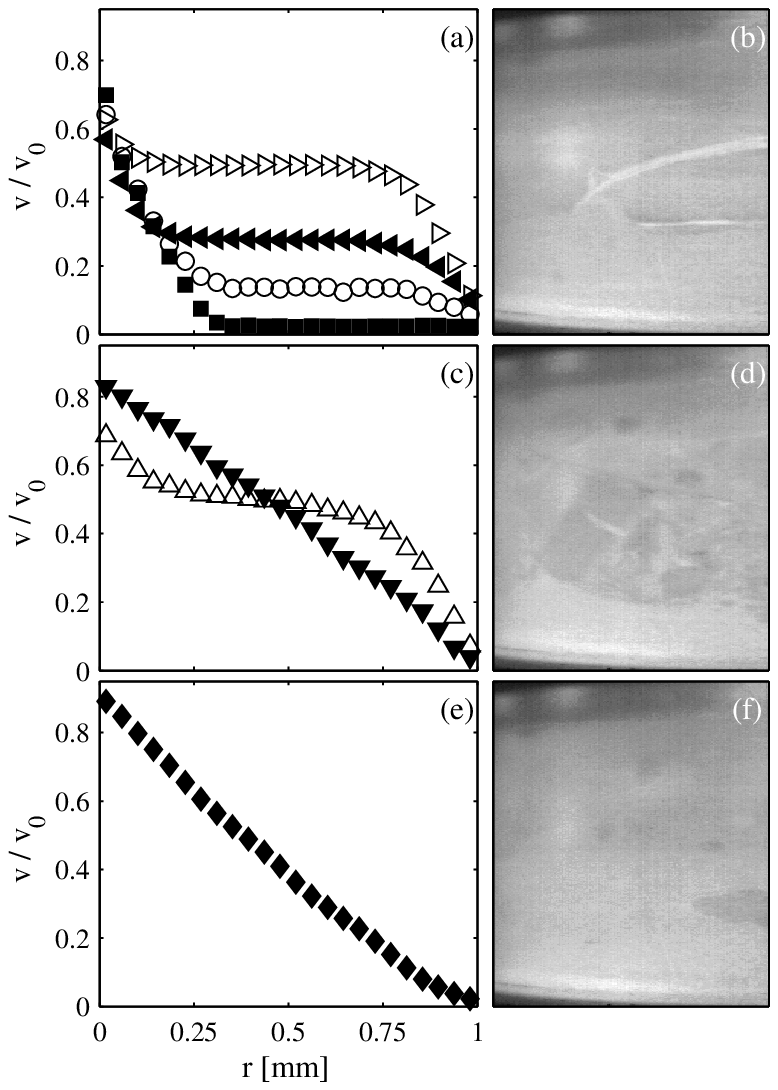}
\caption{Long-time behaviour of a 3~wt.~\% Laponite suspension seeded with 1~wt.~\% microspheres for $\gp=110$~s$^{-1}$. Velocity profiles taken at times (a) $t=1$ ($\blacksquare$), 185 ($\circ$), 222 ($\blacktriangleleft$), and 302~s ($\triangleright$) in regime I, (b) $t=478$ ($\vartriangle$) and 493~s ($\blacktriangledown$) in regime II, and (c) $t=1005$~s ($\blacklozenge$) in regime III. The velocity profiles $v(r)$ have been rescaled by the rotor velocity $v_0$. The right column shows pictures of the sample typical of the three regimes. The height of each picture is 30~mm.}
\label{f.long}
\end{center}
\end{figure}

As shown in Fig.~\ref{f.long}, the long-time behaviour of our sheared samples is {\it not} shear localization. A complex spatio-temporal scenario rather develops on time scales ranging from a few minutes to several hours depending on the applied shear rate. As in Ref.~\cite{Gibaud:2008}, we propose to distinguish three different regimes in the temporal evolution of the flow.

In a first step called regime I (see Fig.~\ref{f.long}(a) and (b)), velocity profiles are seen to evolve from shear localization towards {\it plug-like flow} with strong wall slip within about 5~min for $\gp=110$~s$^{-1}$. Indeed, the arrested region close to the stator ``detaches'' from the fixed wall and expands across the gap, which leads to a velocity profile that is flat (and referred to as ``plug-like'' in the following) over most of the cell except for two small sheared regions of width $\simeq 0.15$~mm at both walls. At the same time, direct observations of the sample indicate that the sample rotates as a solid body and presents large heterogeneities of typical size 1~cm with fracture-like streaks. At the end of regime I, the velocity profiles are symmetrical and the solid material in the center rotates at roughly $v_0/2$, where $v_0$ is the rotor velocity.

\begin{figure}[ht]
\begin{center}
\scalebox{0.9}{\includegraphics{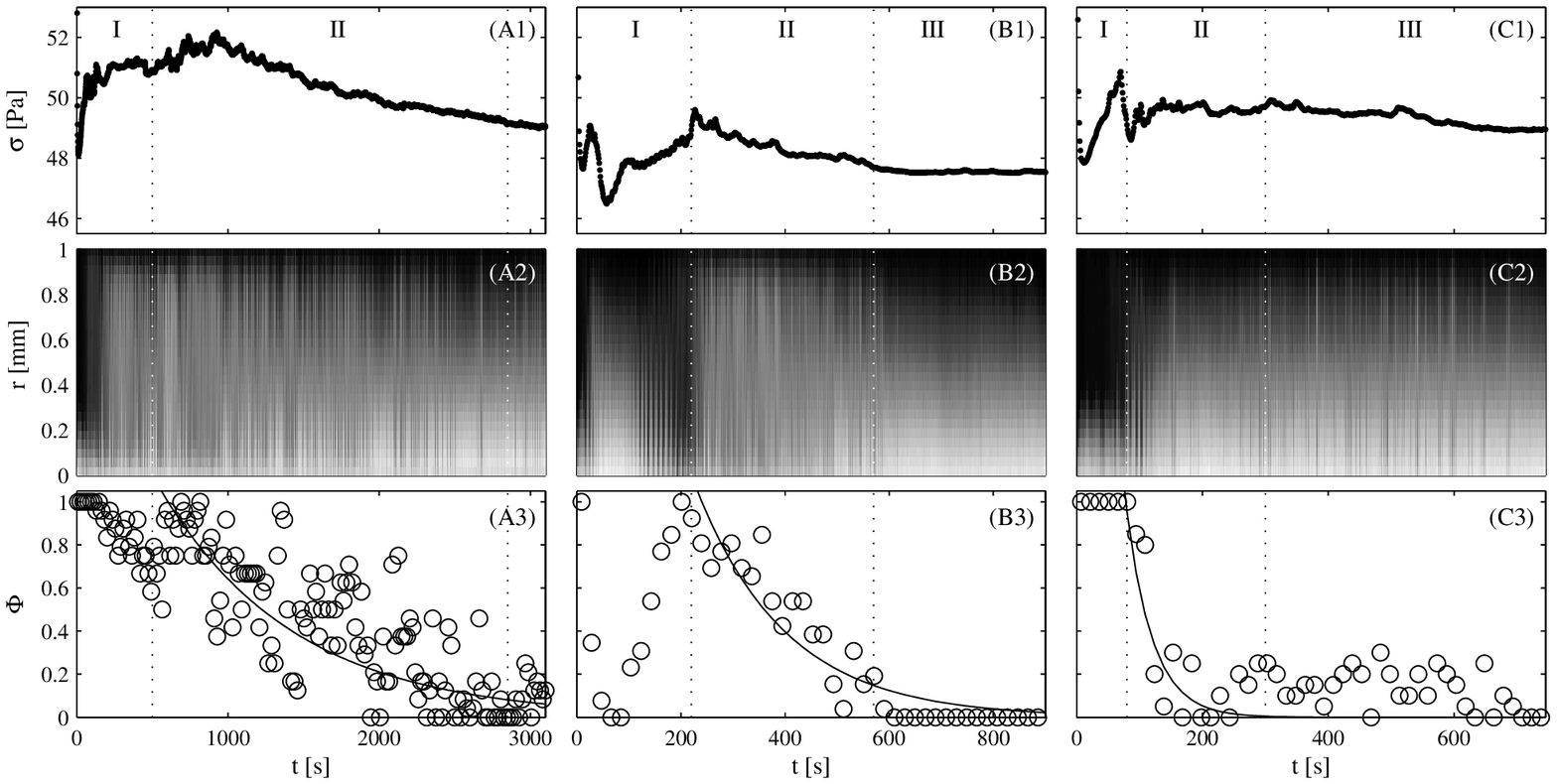}}
\caption{(1) Shear stress response $\sigma(t)$, (2) velocity profiles $v(r,t)$, and (3) fraction $\Phi(t)$ of plug-like velocity profiles in a 3~wt.~\% Laponite suspension seeded with 1~wt.~\% microspheres for (A) $\gp=100$~s$^{-1}$, (B) $\gp=110$~s$^{-1}$, and (C) $\gp=120$~s$^{-1}$. The spatiotemporal diagrams in (2) use linear gray scales ranging from $v=0$ (black) to $v=\gp e$ (white). The solid lines in (3) are exponential relaxations shown to guide the eye across regime II.}
\label{f.sptp}
\end{center}
\end{figure}

\begin{figure}[ht]
\begin{center}
\scalebox{0.9}{\includegraphics{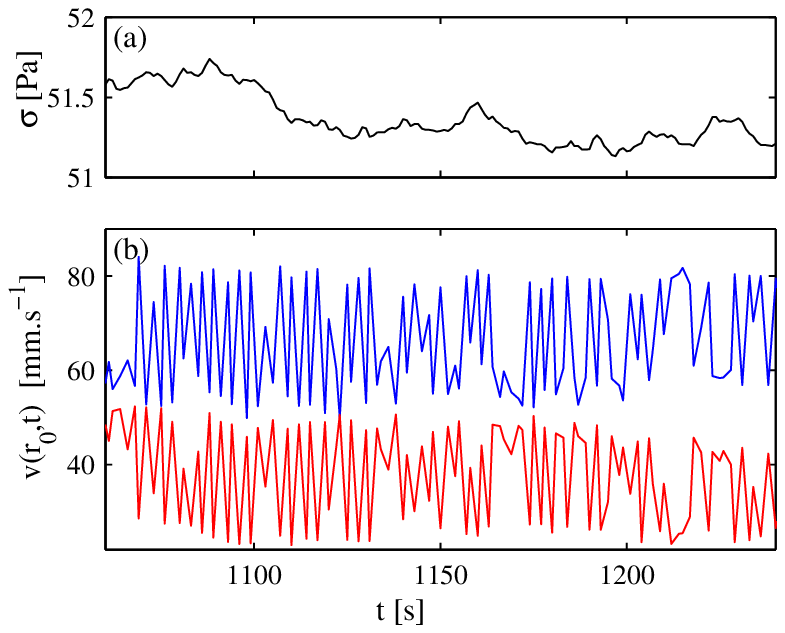}}
\caption{Zoom into regime II for a 3~wt.~\% Laponite suspension seeded with 1~wt.~\% microspheres under $\gp=100$~s$^{-1}$. (a) Shear stress response $\sigma(t)$ and (b) local velocity $v(r_0,t)$ for $r_0=0.1$~mm (blue) and $r_0=0.8$~mm (red).}
\label{f.vit}
\end{center}
\end{figure}

In regime II (see Fig.~\ref{f.long}(c) and (d)), the velocity profiles {\it oscillate} within a few seconds between the plug-like profile described above and a homogeneous Newtonian-like velocity profile with much smaller wall slip (that we shall call a ``linear'' velocity profile below). This is clearly evidenced on the spatiotemporal diagram of $v(r,t)$ in Fig.~\ref{f.sptp}(B2) and on the velocity signals $v(r_0,t)$ shown in Fig.~\ref{f.vit}(b) for two different positions $r_0$ in the gap. Moreover, movies of the sheared samples show that the solid material breaks up into smaller pieces surrounded by fluidized zones in both the azimuthal and vertical directions \cite{Remark:Laponitemovies}. Simultaneous observations and velocity measurements also reveal that linear profiles are recorded whenever a homogeneous fluidized region passes in front of the USV transducer whereas plug-like profiles correspond to solid pieces floating in the sample \cite{Gibaud:2008}.

Therefore, the temporal oscillations of the velocity profiles are simply a consequence of the breaking-up of the sample into a very heterogeneous pattern of smaller blocks that rotate as solid bodies within the fluidized material. Since the time required for recording a single velocity profile is much smaller than the rotation period of the solid blocks, the proportion $\Phi(t)$ of plug-like flow profiles recorded during a given time window centered around time $t$ can be directly interpreted as the proportion of solid pieces within the sample. Yet, this obviously assumes that the time evolution is the same all along the vertical direction, which is not truly the case. In particular, it is often observed that the top and bottom of the Couette cell get fluidized more rapidly than the center of the cell where our velocity measurements are performed. This can most probably be attributed to end effects and, in the absence of any two-dimensional information on the velocity field, we shall neglect the dependence of the flow along the vertical direction. In practice, we discriminate between ``plug-like'' and ``linear'' profiles by monitoring the local shear rate in the middle of the gap: a velocity profile is called plug-like when $\gp(r=0.5)<\gp/2$. $\Phi(t)$ is then defined as the fraction of plug-like profiles over a time window centered at $t$ and of duration ranging from 10 to 50~s depending on the applied shear rate.

In most cases, $\Phi(t)$ decreases slowly throughout regime II, which we interpret as an {\it erosion} of the solid pieces by the fluidized material through viscous stresses. This is directly confirmed by optical imaging that shows a simultaneous decrease of the average size of the solid blocks \cite{Remark:Laponitemovies}, although we were not able to extract more quantitative information about the evolution of solid content due to the low contrast of the images. In the case of Fig.~\ref{f.long}, the erosion process takes about 400~s. Figure~\ref{f.sptp}(B) gathers the full data set for $\gp=110$~s$^{-1}$ by showing the shear stress response $\sigma(t)$ recorded by the rheometer, a spatiotemporal diagram of the velocity profiles $v(r,t)$, and $\Phi(t)$. In particular, Fig.~\ref{f.sptp}(B2) reveals that the temporal variations of the local velocity get slower and slower as $\Phi(t)$ decreases and progressively give way to a homogeneous stationary flow, which corroborates our picture of solid blocks that get smaller and smaller due to erosion. Note that such erosion mostly occurs in the azimuthal and vertical directions since velocity profiles indicate that the radial size of the solid pieces remains comparable to the gap width all along regime II.

Regime II ends when the sample is fully fluidized and only linear velocity profiles are recorded ($\Phi=0$). For longer times, the sample remains fluid and no shear localization is observed in the steady state called regime III (see Fig.~\ref{f.long}(e) and (f)). Slip velocities are always very small in regime III.

In summary, regime I corresponds to a transition from shear localization to plug-like flow at $v_0/2$. In regime II, the solid region that occupies a large part of the gap breaks up into solid blocks that get slowly eroded. In regime III, the sample has finally reached its steady-state characterized by almost linear velocity profiles with negligible wall slip.

\subsubsection{Influence of the applied shear rate}
\label{s.results.sr}
An important issue concerns the robustness of the fragmentation and erosion scenario described above. Here, we explore the influence of the applied shear rate. Figure~\ref{f.sptp} shows the results obtained for $\gp=100$, 110, and 120~s$^{-1}$. In all cases, regimes I, II, and III can always be found and are indicated using dashed lines. The most striking effect of a rather small increase of $\gp$ is the very strong speeding up of the whole process. While it takes about an hour to fully fluidize the sample for $\gp=100$~s$^{-1}$, an almost homogeneous flow is recovered within about 2~min for $\gp=120$~s$^{-1}$. We shall discuss this feature in more details in Section~\ref{s.discuss}.

Moreover, the stress response $\sigma(t)$ may be analyzed in view of the local measurements. Indeed, regime I seems to be associated with rather slow stress variations with large amplitudes of about 2~Pa. This is most clearly seen in Fig.~\ref{f.sptp}(B1) and (C1), although no systematic behaviour of $\sigma(t)$ can be extracted. On the other hand, in regime II, $\sigma(t)$ always fluctuates much more rapidly and with a much smaller amplitude around an average value that slowly decreases. We presume that these noisy features in $\sigma(t)$ are the signature of the local fragmentation and erosion process. However, due to the highly heterogeneous nature of the flow in regime II, most of the local features average out in the global stress response, which therefore presents only small fluctuations around a mean value close to the yield stress $\sigma_y=49\pm 2$~Pa. This is shown in more detail in Fig.~\ref{f.vit} where no clear correlation between the stress response and the local velocity can be detected. Finally, when the steady state is reached in regime III, the shear stress presents a constant value with negligible fluctuations.

\subsubsection{Influence of the microsphere concentration}
\label{s.results.conc}
\begin{figure}[ht]
\begin{center}
\includegraphics{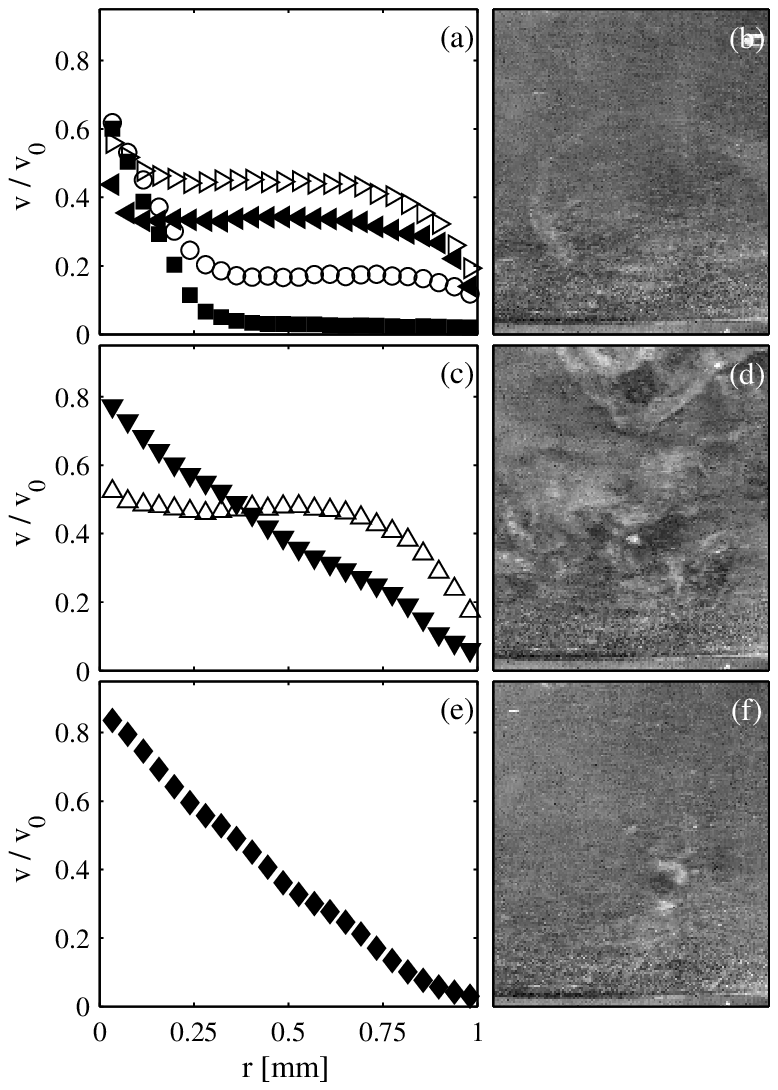}
\caption{Long-time behaviour of a 3~wt.~\% Laponite suspension seeded with 0.3~wt.~\% microspheres for $\gp=39$~s$^{-1}$. Velocity profiles taken at times (a) $t=14$ ($\blacksquare$), 178 ($\circ$), 955 ($\blacktriangleleft$), and 1399~s ($\triangleright$) in regime I, (b) $t=1933$ ($\vartriangle$) and 1974~s ($\blacktriangledown$) in regime II, and (c) $t=2857$~s ($\blacklozenge$) in regime III. The velocity profiles $v(r)$ have been rescaled by the rotor velocity $v_0$. The right column shows pictures of the sample typical of the three regimes. The height of each picture is 30~mm.}
\label{f.long_bis}
\end{center}
\end{figure}

\begin{figure}[ht]
\begin{center}
\scalebox{0.9}{\includegraphics{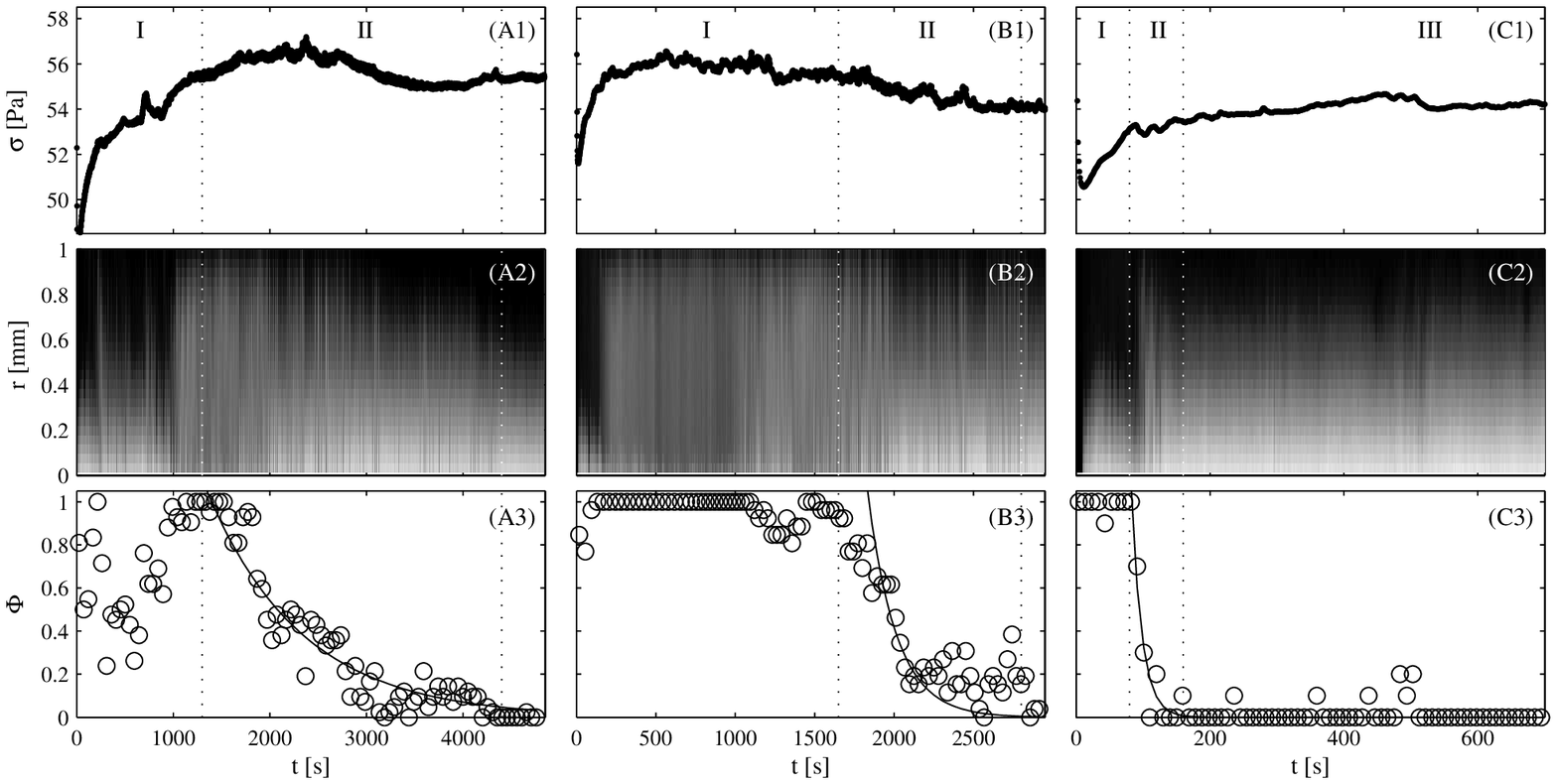}}
\caption{(1) Shear stress response $\sigma(t)$, (2) velocity profiles $v(r,t)$, and (3) fraction $\Phi(t)$ of plug-like velocity profiles in a 3~wt.~\% Laponite suspension seeded with 0.3~wt.~\% microspheres for (A) $\gp=29$~s$^{-1}$, (B) $\gp=39$~s$^{-1}$, and (C) $\gp=50$~s$^{-1}$. The spatiotemporal diagrams in (2) use linear gray scales ranging from $v=0$ (black) to $v=\gp e$ (white). The solid lines in (3) are exponential relaxations shown to guide the eye across regime II.}
\label{f.sptp_bis}
\end{center}
\end{figure}

\begin{figure}[htbp]
\begin{center}
\scalebox{0.9}{\includegraphics{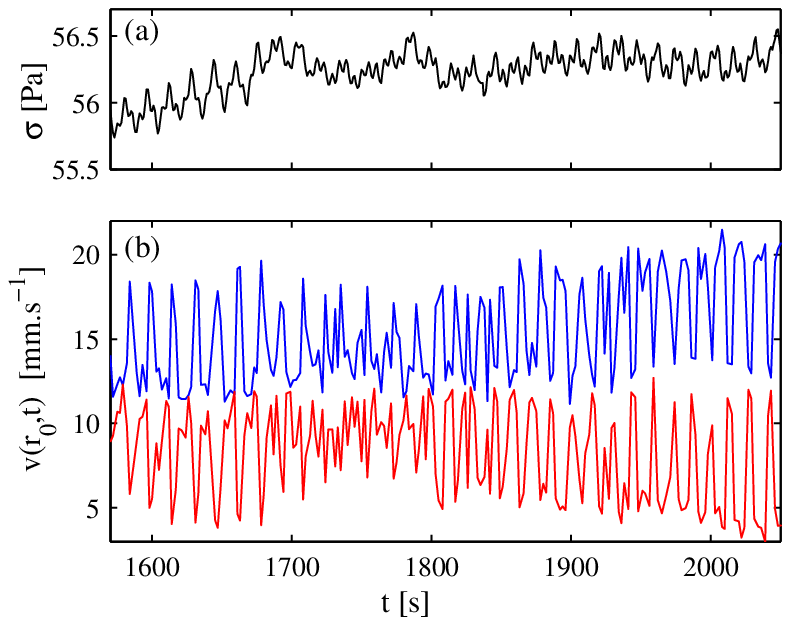}}
\caption{Zoom into regime II for a 3~wt.~\% Laponite suspension seeded with 0.3~wt.~\% microspheres under $\gp=29$~s$^{-1}$. (a) Shear stress response $\sigma(t)$ and (b) local velocity $v(r_0,t)$ for $r_0=0.1$~mm (blue) and $r_0=0.8$~mm (red).}
\label{f.vit_bis}
\end{center}
\end{figure}

The fact that our samples contain glass microspheres in a rather large amount raises legitimate questions about the influence of such seeding particles on our results. To investigate the influence of the microsphere concentration, experiments were repeated with a smaller amount of hollow glass spheres of 0.3~wt.~\%. Figures~\ref{f.long_bis}, \ref{f.sptp_bis}, and \ref{f.vit_bis} show that the flow behaviour unveiled above is also observed for 0.3~wt.~\% microspheres, as reported in Ref.~\cite{Gibaud:2008} where only one applied shear rate ($\gp=17$~s$^{-1}$) was investigated. In particular, Figs.~\ref{f.long} and \ref{f.long_bis} are strikingly similar. The only differences lie (i) in the lower contrast of the images when only 0.3~wt.~\% microspheres are used and (ii) slightly larger slip velocities in the case of the lower microsphere concentration. 

The comparison between Figs.~\ref{f.sptp} and \ref{f.sptp_bis} also proves the {\it robustness} of the fragmentation and erosion scenario. Stress responses display broadly the same features (except for the highest applied shear rate in Fig.~\ref{f.sptp_bis} for which an increasing $\sigma(t)$ was recorded). Both the qualitative evolutions of $v(r,t)$ and of $\Phi(t)$ do not seem to depend on the microsphere concentration.

However, the values of the shear rate at which this complex yielding behaviour is observed are noticeably different. Indeed, for 1~wt.~\% microspheres, the fragmentation and erosion process is observed for $\gp\simeq 85$--250~s$^{-1}$ whereas this range is $\gp\simeq 15$--65~s$^{-1}$ for 0.3~wt.~\% microspheres. This remarkable effect of the microsphere concentration is further discussed in the next Section. Moreover, as revealed by Fig.~\ref{f.vit_bis} for a Laponite suspension seeded with 0.3~wt.~\% microspheres and sheared at $\gp=29$~s$^{-1}$, oscillations may be detected in the stress response with the same period as the velocity oscillations. Although the relative stress fluctuations remain small, such a correlation hints at a less heterogeneous flow pattern and  possibly more collective fragmentation and erosion processes.

\section{Discussion and model}
\label{s.discuss}

\subsection{Comparison with other works}
The data presented in this paper show that the fragmentation and erosion scenario first reported in Ref.~\cite{Gibaud:2008} is found over a large range of shear rates and is robust against a change of microsphere concentration. To our knowledge, only two other works on SGM close to yielding may have reported behaviours similar to the one presented here. First, intermittent jammed states were observed together with the occurence of strong wall slip in a glassy colloidal star polymer using NMR velocimetry \cite{Holmes:2004}. Second, more recently, oscillating velocities were reported in 3~wt.~\% Laponite samples under low shear rates \cite{Ianni:2008}. In the latter case, the two extreme velocity profiles were reconstructed through pointwise DLS measurements. Figure~4 in Ref.~\cite{Ianni:2008} is broadly consistent with our observations, e.g., with our Figs.~\ref{f.long}(c) and \ref{f.long_bis}(c). Although the oscillating velocity profiles were interpreted in terms of a stick-slip instability by the authors, we believe that this behaviour may actually be due to fragmentation and fluidization as described here. Unfortunately, in Ref.~\cite{Ianni:2008}, the evolution of the sample on very long time scales was not investigated and optical imaging was not possible since the sample did not contain any seeding particles and was thus perfectly transparent \cite{Ianni:PrivateComm}.

\subsection{Characteristic time scale for erosion}
\begin{figure}[htbp]
\begin{center}
\includegraphics{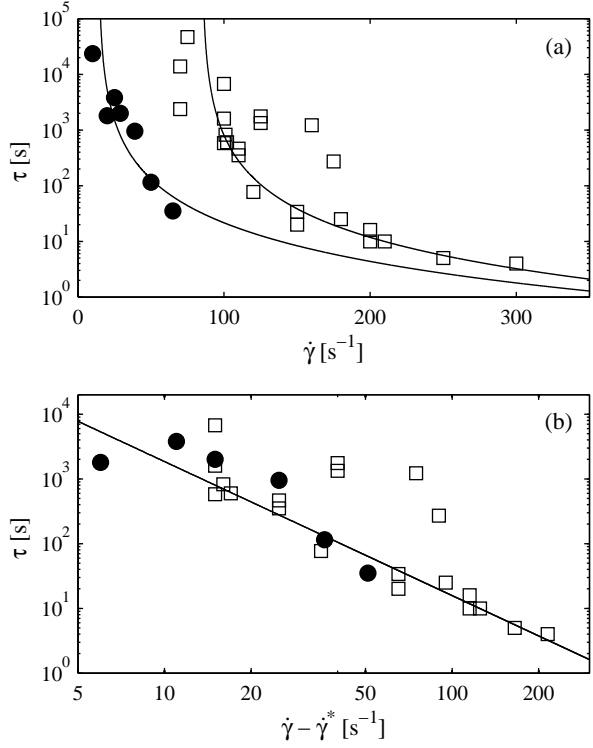}
\caption{Characteristic erosion time $\tau$ in regime II vs (a) the applied shear rate $\gp$ and (b) $\gp-\gps$ for 3~wt.~\% Laponite suspensions seeded with 1~wt.~\% ($\square$) and 0.3~wt.~\% ($\bullet$) microspheres. The solid lines are $\tau=A(\gp-\gps)^{-n}$ with $n=2.1$, $A=2.2\;10^5$, and $\gps=85$~s$^{-1}$ ($\gps=14$~s$^{-1}$ resp.) for the 1~wt.~\% (0.3~wt.~\% resp.) microsphere concentration.}
\label{f.tau}
\end{center}
\end{figure}

To further discuss these results, we extracted a characteristic time $\tau$ for the erosion process defined as the time it takes for $\Phi(t)$ to relax from 0.9 to 0.1 during regime II (see also the solid lines in Figs.~\ref{f.sptp} and \ref{f.sptp_bis}). In Fig.~\ref{f.tau}(a), this characteristic time is plotted as a function of the applied shear rate for the two microsphere concentrations investigated here. It clearly appears that $\tau$ varies by more than three orders of magnitude within a rather small range of shear rates of order 20~s$^{-1}$. The time scale for erosion seems to diverge for shear rates smaller than $\gps\simeq 85$~s$^{-1}$ for 1~wt.~\% microspheres and $\gps\simeq 15$~s$^{-1}$ for 0.3~wt.~\% microspheres. In fact, for $\gp\lesssim\gps$, only plug-like flow profiles were observed over the longest experimental waiting times of about $10^5$~s, so that fragmentation actually never occured. For the largest shear rates, the material becomes fluid within seconds and, due to our limited time resolution of about 1~s per velocity profile, relaxation times shorter than a few seconds cannot be accessed.

The large scatter of these data may be attributed to poor reproducibility of the yielding behaviour. Indeed, experiments performed on two different samples from the same batch at the same shear rate may yield estimates for $\tau$ that differ by almost a decade. Uncertainty also arises from the difficulty in precisely locating the beginning and the end of regime II. As seen for instance in Figs.~\ref{f.sptp}(A) and \ref{f.sptp_bis}(B), $\Phi(t)$ may start to decrease and go back to $\Phi=1$ within a few minutes so that fragmentation seems to begins but soon stalls before starting again. Figure~\ref{f.sptp}(C) also shows that very small solid pieces may persist for very long times even though $\Phi(t)$ drops to values of order 0.2 in less than 10~s. Such behaviours, which could also be ascribed to spatiotemporal heterogeneities of the flow in the vertical direction, lead to large error bars on $\tau$.

Still, the very steep decrease of $\tau$ for $\gp\gtrsim\gps$ prompts us to model the experimental data by the phenomenological law:
\begin{equation}
\tau=A(\gp-\gps)^{-n}\,,
\label{e.tau}
\end{equation}
where $n$ is some characteristic positive exponent. Due to the large spread of the data, fits to equation (\ref{e.tau}) with three free parameters do not yield reliable results. However, forcing the value $\gps=85$~s$^{-1}$ and excluding the most deviant data points leads to a satisfactory fit with $n=2.1$ and $A=2.2\;10^5$ for 1~wt.~\% microspheres (see solid line in Fig.~\ref{f.tau}(a)). Note that when $\gps$ is varied from 75 to 95~s$^{-1}$, the exponent $n$ varies from 2.5 to 1.5. Therefore, in order to account for the large experimental scatter, we shall conclude that $\gps=85\pm 10$~s$^{-1}$ and $n=2.1\pm 0.5$ provide a good modelling of the 1~wt.~\% microsphere data.

In Fig.~\ref{f.tau}(b), $\tau$ is plotted in logarithmic scales against $\gp-\gps$ using $\gps=85$~s$^{-1}$ for 1~wt.~\% microspheres and $\gps=14$~s$^{-1}$ for 0.3~wt.~\% microspheres respectively. The rather good collapse of both data sets allows us to conclude that $n$ and $A$ do not significantly depend on the microsphere concentration while the two values for $\gps$ are clearly different. 

\subsection{A simple model for erosion}
\label{s.discuss.model}
In order to model the erosion process, we suppose that the sample at the beginning of regime II is characterized by some distribution of solid fragments of initial size $R_0$. To account for the progressive fragmentation of the sample at the end of regime I, each fragment is assumed to start being eroded at an initial time $t_0$ characterized by a given probability distribution.

Erosion results from viscous friction at the surface of the solid pieces. Since the solid pieces have a yield stress $\sigma_y$, the viscous stress available for erosion is $\eta_f\gp-\sigma_y$, where $\eta_f$ is the viscosity of the surrounding fluid, here, the fluidized Laponite suspension. Rewriting this stress as $\eta_f(\gp-\gps)$, with $\gps=\sigma_y/\eta_f$, shows that the characteristic time scale for erosion is $1/(\gp-\gps)$.

We then introduce the size $r_b$ of the ``elementary brick'' that detaches from solid fragments due to erosion. Since erosion occurs at the surface, we expect the number of bricks $N_v=R(t)^3/r_b^3$ in a fragment of size $R(t)$ at time $t$ to be linked to the number of bricks at the surface $N_s=R(t)^2/r_b^2$ by:
\begin{equation}
\frac{\dd N_v}{\dd t}\propto -(\gp-\gps) N_s\,.
\end{equation}
This simple dimensional argument leads to: 
\begin{equation}
\frac{\dd R}{\dd t}= -(\gp-\gps) r_b\,,
\label{e.R}
\end{equation}
up to some dimensionless multiplicative constant of order unity which we may incorporate into $r_b$. Note that the same equation for $R(t)$ would be found if one rather considers a two-dimensional problem where the solid blocks are discs of constant thickness, which may be more relevant for the present experiments.

Finally, we compute $\Phi(t)$ from the distribution of fragment sizes $\{R_j(t)\}_{j=1\dots N}$ at time $t$ as:
\begin{equation}
\Phi(t)=\frac{\sum_{j=1}^N R_j(t)}{\sum_{j=1}^N R_{j0}}\,,
\label{e.phi}
\end{equation}
where $N$ is the total number of fragments and $R_{j0}$ is the initial size of fragment number $j$ whose erosion starts at time $t_{j0}$. From Eq.~(\ref{e.R}), one has:
\begin{equation}
R_j(t)=\left\{ \begin{array}{l}
	R_{j0} \hbox{~if~} t\le t_{j0}\\
	0 \hbox{~for~} t\ge t_{j0}+\tau_j\\
	R_{j0}-r_b(\gp-\gps)(t-t_{j0}) \hbox{~otherwise.}\end{array} \right.
\label{e.Rj}
\end{equation}
where $\tau_j=\frac{R_{j0}}{r_b}\,\frac{1}{\gp-\gps}$ is the total erosion time for fragment number $j$.

The definition of $\Phi(t)$ in Eq.~(\ref{e.phi}) was chosen in order to be consistent with the experimental measurements of $\Phi(t)$ presented in Section~\ref{s.usv}. Indeed, these measurements, based on the proportion of plug-like velocity profiles measured within a given time window, correspond to one-dimensional measurements and should be related to the linear size $R_j(t)$ of the solid pieces. Furthermore, although the exact shape of $\Phi(t)$ depends on the details of the distributions chosen for $R_0$ and $t_0$, it is clear from Eq.~(\ref{e.Rj}) that the characteristic time scale for $\Phi(t)$ should be of the order of:
\begin{equation}
\tau\simeq\frac{\overline{R_0}}{r_b}\,\frac{1}{\gp-\gps}\,,
\label{e.tau_model}
\end{equation}
where $\overline{R_0}$ is the average initial fragment size.

Comparing with Eq.~(\ref{e.tau}), our simple model predicts a $n=1$ scaling instead of the experimental exponent $n=2.1\pm 0.5$. This is most probably due to the crude assumptions (i) that the distributions for $R_0$ and $t_0$ (in particular $\overline{R_0}$) do not depend on the applied shear rate, (ii) that erosion is governed by the detachment of elementary bricks whose size is independent of both time and shear rate, and (iii) that there is a direct link between the yield stress of the bulk material at rest and the critical shear rate $\gps$. Still, the model predicts a rather good order of magnitude for $\tau$. Indeed, if $r_b$ is estimated from the elastic modulus $G_0$ as $r_b=(k_B T/G_0)^{1/3}$ with $G_0=500$~Pa, one finds $r_b\simeq 40$~nm, which roughly corresponds to the diameter of the Laponite discs. Taking $\overline{R_0}\sim 1$~cm for the initial size of the fragments as observed in the experiments and $\gp-\gps=10$~s$^{-1}$, Eq.(\ref{e.tau_model}) leads to $\tau\sim 10^4$~s in statisfactory agreement with the measurements reported in Fig.~\ref{f.tau}.

In order to illustrate our simple model, we chose to focus on the case where $R_{j0}$ is fixed to a given value $R_0$ for all solid fragments and $t_{j0}$ is characterized by a Gaussian distribution of mean $\overline{t_0}$ and standard deviation $\delta t_0/\sqrt{2}$. In this case, for $N\gg 1$, Eqs.~(\ref{e.phi}) and (\ref{e.Rj}) become respectively:
\begin{equation}
\Phi(t)=\int_{-\infty}^\infty \frac{R(t,t_0)}{R_0}\,p(t_0)\,\dd t_0\,,
\label{e.phi2}
\end{equation}
where $p(t_0)=\exp(-(t_0-\overline{t_0})^2/\delta t_0^2)/(\sqrt{\pi}\delta t_0)$ is the probability density function of the initial time $t_0$, and 
\begin{equation}
\frac{R(t,t_0)}{R_0}=\left\{ \begin{array}{ll}
	0 &\hbox{~if~} t_0\le t-\tau\\
	1-\frac{t-t_0}{\tau} &\hbox{~if~} t-\tau<t_0<t\\
	1 &\hbox{~if~} t_0\ge t\,, \end{array} \right.
\label{e.R2}
\end{equation}
where we have noted $\tau=\frac{R_{0}}{r_b}\,\frac{1}{\gp-\gps}$ the erosion time for a single fragment. Inserting Eq.~(\ref{e.R2}) into Eq.~(\ref{e.phi2}) and using the dimensionless time $\widetilde{t}=(t-\overline{t_0})/\delta t_0$ leads to the following analytical expression:
\begin{equation}
\Phi(\widetilde{t})=\frac{1-\hbox{\rm erf}(\widetilde{t}-\alpha)}{2}+\int_{\widetilde{t}-\alpha}^{\widetilde{t}}\left(1-\frac{\widetilde{t}-y}{\alpha}\right)\frac{e^{-y^2}}{\sqrt{\pi}}\,\dd y\,,
\label{e.phi3}
\end{equation}
where we have defined the ratio of the erosion time to the standard deviation of $t_0$ as $\alpha=\tau/\delta t_0$ and ``erf'' stands for the error function: $\hbox{erf}(x)=\frac{2}{\sqrt{\pi}}\int_0^x e^{-y^2}\,\dd y$.

\begin{figure}[ht]
\begin{center}
\includegraphics{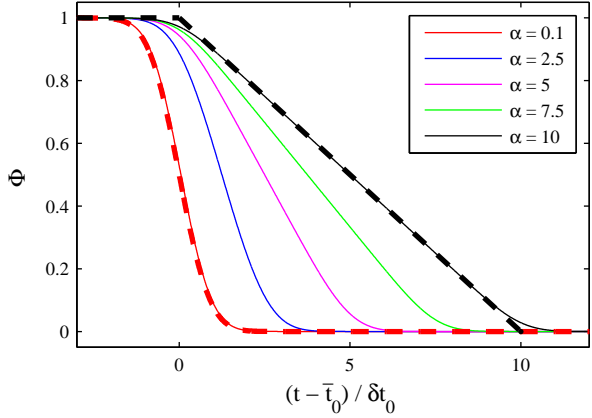}
\caption{Fraction of solid fragments $\Phi$ computed from Eq.~(\ref{e.phi3}) as function of the reduced time $\widetilde{t}=(t-\overline{t_0})/\delta t_0$ for different values of $\alpha=\tau/\delta t_0$ (see text). The red and black dashed lines are respectively $\Phi(\widetilde{t})=(1-\hbox{\rm erf}(\widetilde{t}))/2$ and $\Phi(\widetilde{t})=1-\widetilde{t}/\alpha$ for $0<\widetilde{t}<\alpha=10$.}
\label{f.model1}
\end{center}
\end{figure}

Figure~\ref{f.model1} shows the predicted $\Phi(\widetilde{t})$ for various values of $\alpha$. In the limit $\alpha\ll 1$, where the distribution of initial times is very wide compared to the erosion time, Eq.~(\ref{e.phi3}) reduces to $\Phi(\widetilde{t})=(1-\hbox{\rm erf}(\widetilde{t}))/2$ as shown by the red dashed line. On the other hand, the opposite limit $\alpha\gg 1$ corresponds to very slow erosion so that the initial times can all be considered as equal to $\overline{t_0}$ and the dynamics of $\Phi(t)$ is simply that of a single fragment, i.e. a linearly decreasing function (see black dashed line).

\begin{figure}[ht]
\begin{center}
\includegraphics{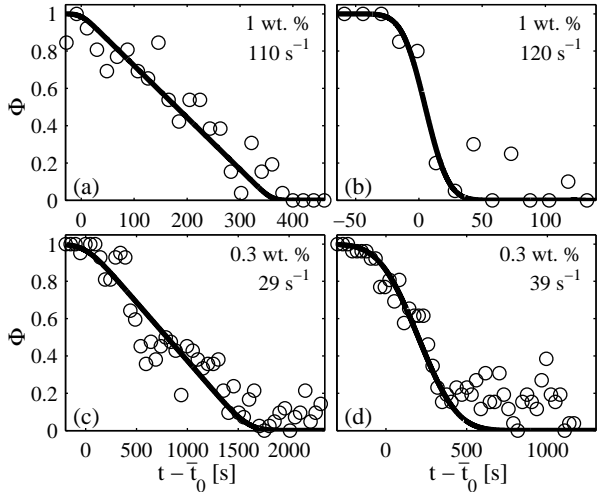}
\caption{Predictions of Eq.~(\ref{e.phi3}) (thick solid lines) compared to experimental data ($\circ$). (a) $\overline{t_0}=210$~s, $\delta t_0=20$~s, and $\tau=360$~s ($\alpha=18$). (b) $\overline{t_0}=110$~s, $\delta t_0=20$~s, and $\tau=10$~s ($\alpha=0.5$). (c) $\overline{t_0}=1430$~s, $\delta t_0=200$~s, and $\tau=1600$~s ($\alpha=8$). (d) $\overline{t_0}=1750$~s, $\delta t_0=200$~s, and $\tau=400$~s ($\alpha=2$). The experimental conditions, namely the microsphere weight fraction and the applied shear rate, are indicated directly on the figures.}
\label{f.model2}
\end{center}
\end{figure}

Finally, a comparison between the predictions of Eq.~(\ref{e.phi3}) and experimental $\Phi(t)$ data taken from Figs.~\ref{f.sptp} and \ref{f.sptp_bis} is shown in Fig.~\ref{f.model2}. In the model, the average time $\overline{t_1}$ is directly estimated from the experimental data by looking for the time at which regime II begins. For a given microsphere concentration, good agreement could be found for two different applied shear rates by using the same standard deviation $\delta t_0$ and varying only the erosion time $\tau$. The transition from linear to curved shapes for $\Phi(t)$ as the erosion time scale is decreased, i.e. as the shear rate is increased, constitutes the most prominent feature of our simple approach. In spite of large experimental scatter, this feature is noticeable at least in Figs.~\ref{f.model2}(a) and (d). However, as already noted, the model is not compatible with the strong divergence of $\tau$ as $1/(\gp-\gps)^2$. Within the erosion model proposed above, the experimental scaling thus points to an additional dependence of the initial fragment size $R_0$ on $\gp$ as $1/(\gp-\gps)$.

\subsection{Open questions}

The main open issue raised by our results concerns the mechanism for fragmentation and its relation to boundary conditions. As reported in Ref.~\cite{Gibaud:2008}, fragmentation is only observed with ``smooth'' surfaces and must be somehow linked to boundary conditions. We proposed to explain the observed fragmentation by temporary sticking events localized at specific sites where the yield stress may be overcome, giving rise to a fluidized region separating solid pieces \cite{Gibaud:2008}. Note that such events are highly probable because the wall roughness (about 15~nm for our ``smooth'' cell) is comparable to the size of the Laponite platelets.

However, the fact that the characteristic shear rate $\gps$ strongly depends on the amount of seeding microspheres calls for a more subtle interpretation. One may even wonder whether the fragmentation and erosion scenario would still be observed if the microsphere concentration was further reduced or if the Laponite suspension did not contain any microsphere at all. The observation of oscillatory behaviours in pure Laponite \cite{Ianni:2008,Ianni:PrivateComm} tends to support the possibility of our scenario even in the absence of any seeding particle. In our experiments, we lowered the microsphere concentration down to 0.1~wt.~\%, which still allowed us to measure velocity profiles with reasonable accuracy. Fragmentation and erosion could be observed in our ``smooth'' cell only when wall slip was promoted in an even better way by lubricating the cell walls using vacuum grease (data not shown). We also tried to seed our Laponite suspensions with polystyrene (rather than glass) microspheres. This led to the same observation that the fragmentation and erosion scenario needed very slippery boundary conditions to develop. 
 
Therefore, we may speculate that surface roughness is not the only relevant control parameter but that microspheres--Laponite and/or microspheres--surface interactions may also play a significant role and influence the values of $\gps$ and $\tau$. However, a systematic study of these effects, as well as those of sample age and Laponite concentration, still remains to be performed.

Finally, other important open questions concern (i) the link between the critical shear rate $\gps$ for erosion and the characteristic shear rate $\gpc$ involved in shear localization at early times or in the absence of wall slip and (ii) the possible effects of confinement on the observed shear-induced fragmentation.

\section{Conclusion}
\label{s.conclu}
In this work, we have investigated in details the original yielding scenario of Laponite suspensions seeded with glass microspheres first reported in Ref.~\cite{Gibaud:2008} and triggered by slippery boundary conditions. Such a scenario involves shear localization at short times followed by a slow evolution towards plug-like flow (regime I). Then, over much longer time scales, fragmentation and erosion are observed provided the applied shear rate is larger than some critical value $\gps$. Fragmentation leads to apparent oscillations in the velocity profiles (regime II). Except for small noisy and/or oscillatory features in the stress response, the very large local fluctuations of the flow field have no significant impact on the global rheological response. Depending on the shear rate, a homogeneous flow (regime III) is recovered within a few minutes up to several hours. This scenario was reported for a wide range of shear rates and for two different microsphere concentrations. The characteristic time scale for erosion roughly follows a $1/(\gp-\gps)^2$ power law, where $\gps$ depends on the microsphere concentration. A simple model was introduced to qualitatively reproduce the erosion process. Yet, a better understanding and characterization of the fragmentation and erosion phenomena in such a soft glassy material are still needed both from the experimental and from the theoretical sides. These results also prompt us to look for similar yielding behaviours in other materials under slippery conditions and on long time scales.

\vspace{0.5cm}
The authors wish to thank P.~Jop and L.~Petit for their help on the system as well as A.~Piednoir and M.~Monchanin for the atomic force microscopy measurements. They are very grateful to J.-F. Palierne for letting them use his rheometer. Fruitful discussions with L.~Bocquet, A.~Colin, T.~Divoux, W.~Poon, J.-B. Salmon, and C.~Ybert are acknowledged. This work was supported by ANR under project SLLOCDYN.

\clearpage

\clearpage

\clearpage

\clearpage

\end{document}